\input psbox.tex
\splitfile{\jobname}
\let\autojoin=\relax

\documentstyle[12pt]{article}
\begin{document}
\makeatletter
\input psbox.tex
\makeatother
\begin{titlepage}
\hfill{U\'{S}L-TH-93-01}
\vspace{5 mm}

\begin{center}
{\Large {\bf {The neutrino ball model of a quasar.}}}\\
\vspace{7mm}
{\bf R.Ma\'{n}ka\footnote{email: manka@plktus11.bitnet, internet:
155.158.11.6},I.Bednarek, D.Karczewska }\\
\vspace{3mm}
{\sl Department of Astrophysics and Cosmology,}\\
{\sl University of Silesia,Uniwersytecka 4, 40-007 Katowice, Poland}\\
\end{center}
\setcounter{equation}{0}
\vspace{8 mm}
\centerline{ABSTRACT}
\vspace{2 mm}

It is suggested that the nonorthodox model of a quasar   as
 a neutrino ball described in terms of the standard model extended by
adding right-handed neutrinos and the Majorana scalar
field can be presented
in order to explain a quasar
as a body of weak interacting neutrinos. Neutrino interaction with the
scalar Majorana field violates the lepton number and produces the mass
splitting
of neutrino due to the sea-saw mechanism.
In this model a quasar is an object
which appears in the result of the first order cosmological
phase transition.
In this interpretation a quasar may be regarded as a ball filled with
Dirac neutrinos and can be treated as a remnant of phase transition
with unbroken global lepton symmetry.
In this paper we study the macroscopic parameters of such a configuration.
In the result the mass-radius curve $M(R)$ for the quasar is obtained.
\vfill
\leftline{PACS number(s): 98.80.Cq, 12.15.Cc}
\vspace{2 mm}
\end{titlepage}

\section{Introduction.}

The scalar field may have played an important role in the evolution of the
universe \cite{ahg:1981},\cite{adl:1982}.
The question whether the scalar field can play a role of similar importance in
astrophysics seems to be natural.
Models of  galaxy and star formation lead to the question
whether  the gravitational equilibrium configuration of massive scalar fields
may exist. These types of objects made up of scalar fields can be considered
in terms of
nontopological solitons.
\cite{oo:1989}.
Nontopological soliton solution of classical field theory was
introduced by many authors. Up till now properties
of many adopted form of this solutions such as soliton stars \cite{kk:1980}
have been intensively studied.\\
Much more interesting is a fermionic star where bosonic  soliton envelope
is filled with fermions.
The objective here is to examine the other model of quasars  \cite{add:1990}
as a neutrino ball
\cite{bh:1987} which is basing on the
 standard model.
This is a nonorthodox model of a quasar.
Nowadays the most popular model of a quasar is that its energy is
gravitational
in origin and due to accretion of matter on a
supermassive black hole ( M $ \sim 10^8 \ M_{\odot}$) which is situated
inside the quasar. This model explains the nature of the quasar only
on the base of the gravity. In this paper we try to explain the quasar
as the result of the weak neutrinos interaction. The gravity in this model
play only the auxiliary role. In this model the quasar is an object
which appears in the result of the first order cosmological
phase transition connected with the Weinberg-Salam model \cite{lhr:1987},
\cite{pl:1981}.
\section{The theoretical background.}
The Glashow-Weinberg-Salam model with $SU_{L}(2)\times U_{Y}(1)$ symmetry
is described by the Lagrange function
\begin{eqnarray}
{\cal L} & = & -\frac{1}{4}F^{a}_{\mu \nu}F^{a\mu \nu} - B_{\mu \nu}B^{\mu
\nu} + (D_{\mu}H)^{+}D^{\mu}H - U(H) \\  \nonumber
& +& i\bar{L}\gamma^{\mu}D_{\mu}L +i\bar{e}_{R}\gamma^{\mu}D_{\mu}e_{R} +
h(H\bar{L})e_{R}
\end{eqnarray}
with the $SU_{L}(2)$ field strength tensor
\begin{equation}
F^{a}_{\mu \nu} = \partial_{\mu}W^{a}_{\nu} - \partial_{\nu}W^{a}_{\mu}
+g\epsilon_{abc}W^{b}_{\mu}W^{c\nu}
\end{equation}
and the $U_{Y}(1)$ field tensor
\begin{equation}
B_{\mu \nu} = \partial_{\mu}B_{\nu} - \partial_{\nu}B_{\mu}.
\end{equation}
The covariant derivative is given by
\begin{equation}
D_{\mu} = \partial _{\mu} -\frac{1}{2}igW^{\alpha}_{\mu}T^{\alpha}-
\frac{1}{2}g^{'}YB_{\mu}
\end{equation}
$\alpha =0,1,2,3$.
The index $\alpha =0$ is connected with the $U_{Y}(1)$ group and $\alpha =
1,2,3$ with the $SU_{L}(2)$
group.
In this notation
\begin{equation}
W_{\mu}^{\alpha} = (W_{\mu}^0 = B_{\mu}, W_{\mu}^a)
\end{equation}
where $B_{\mu}$ and $W^{a}_{\mu}$ are a local gauge fields associated with the
$U_{Y}(1)$ and $SU_L(2)$ symmetry group, respectively. $Y$ is a hipercharge.
The gauge group is the simply
multiplication of $U_{Y}(1)$ and $SU_{L}(2)$ so there are two gauge couplings
$g$ and $g'$.
Generators of the gauge groups are unit matrix for $U_{Y}(1)$ and Pauli
matrices for $SU_{L}(2)$
\begin{equation}
T^{\alpha} = (T^{0} = I, T^{a} = \sigma^{a})
\end{equation}
Here we also adopted the notation
\begin{equation}
L = \left( \begin{array}{c}
\nu_{L} \\ e_{L}
\end{array}   \right),L=e,\nu,\tau
\end{equation}
In the simplest version of the standard model a doublet of Higgs field is
introduced
\begin{equation}
 H = \left( \begin{array}{c}
H^{+} \\ H^{0}
\end{array}   \right)
\end{equation}
with the Higgs potential
\begin{equation}
U(H^{+},H) = \lambda(H^{+}H - \frac{1}{2}\upsilon^{2})^{2}
\end{equation}
The form of the potential leads to a degeneracy of the vacuum and to a
nonvanishing vacuum expectation value $<H^{0}>_{0}$ of the
Higgs field
\begin{equation}
<H^{0}>_{0} = \frac{1}{\sqrt{2}}\left( \begin{array}{c}
0 \\ \upsilon
\end{array}   \right)
\end{equation}
and in consequence to the fermion and boson masses.
The minimal leptons set was extended beyond the standard model by
adding the right-handed neutrinos which are singlets with respect to the
$SU_L (2)$ group.
An important feature of this model is the presence
of a complex scalar field $\varphi =\frac{1}{\sqrt2}e^{i\chi}(\rho + V)$ where
$\rho$ and $\chi$
are respectively a massive and massless (Goldstone) field.
$\chi$  denotes the singlet majoron field.
The global lepton number conservation will be respect if a lepton number $+2$
is assigned
to $\varphi$.
The Lagrangian of the additional fermion and scalar fields looks as follows
\begin{equation}
{\cal L'} = \partial_{\mu}\varphi^{+}\partial^{\mu}\varphi - U(\varphi)
+i\bar{\nu}_{R}\gamma^{\mu}\partial_{\mu}\nu_{R}
+h_{2}(\bar{L}\epsilon H)\nu_{R} + h_{3}(\bar{\nu}_{R}^{c}\nu_{R}\varphi +
h.c.)
\end{equation}
where
\begin{equation}
U(\varphi) = \lambda'\varphi^{2}(\varphi-\varphi_{0})^{2}
\end{equation}
{}From the form of the Lagrangian it is evident that the fermion-scalar fields
interaction is described by Yukawa-type couplings.  Since we are interested
only
in the neutrino masses we described a model that embodies this fact.\\
 The Higgs mechanism generates not only the Dirac mass
\begin{equation}
m_{D\nu} = \frac{1}{\sqrt{2}}h_{2}\upsilon
\end{equation}
but the lepton number violating Majorana mass
\begin{equation}
M = \frac{1}{\sqrt{2}}h_{3}V
\end{equation}
as well.
Thus the neutrino mass matrix can be written as follows
  $$
{\cal M} =
\pmatrix{0 & m_{D\nu} \cr m_{D\nu} \eta & M }
$$
In the case $M=0$ only the Dirac neutrino may be obtained.
In general, it should have the same mass as the electron or quark
($ \sim 1\  MeV$). In the broken phase  due to the see-saw
mechanism we obtain two Majorana mass eigenstates \cite{bb:1992}
 \begin{equation}
m_{\nu,M} =\frac{ 1}{2}M \{1 \mp \sqrt{1+4 {h_{2}}^2 ({\frac{
\upsilon}{M}})^2} \}
\sim \{ -\frac{{m_{\nu,D}}^2}{M} , M\}
 \end{equation}
So, for $m_{\nu,D} \sim 1 MeV$ and $ M \sim 1 $ TeV we have
$ m_{1,\nu,M} \sim 1$ eV and $ m_{2,\nu,M} \sim  1$ TeV for the
Majorana neutrinos.
\section{The soliton model of the neutrino ball}
In order to conclude if a neutrino ball could be consider as a possible,
alternative model of a quasar it is essential to
estimate the most important parameters
 of the neutrino ball. Quantities of primary interest to us are
the critical radius, mass and density.
In the case of degenerate vacuum in the presence of fermions the total
energy of the soliton is a sum of the fermion energy ${\cal E}_{f}$ and
the surface energy ${\cal E}_{\sigma}$ and is equal to
\begin{equation}
{\cal E}={\cal E}_f + 4 \pi \int^{\infty}_{0} r^2 dr \{|\frac{ d\varphi}{dr
}|^2
+U(\varphi) \}
\end{equation}
The fermion fields are treated as a degenerate relativistic Fermi gas.
Because of the very large number of fermions the Thomas-Fermi approximation
was adopted to calculate the contribution of the fermions.
The fermion energy ${\cal E}_f$ corresponds to the  repulsive  force
coming from the
Pauli principle
  \begin{equation}
{\cal E}_f = \frac{4V}{(2\pi)^{3} }  \pi k^4_F
 \end{equation}
 where
  \begin{equation}
  k_F = A^{\frac{ 1}{3 }} \frac{ N^{\frac{ 1}{3 }}}{R }
   \end{equation}
is the Fermi momentum.
It defines also the neutrino number
 \begin{equation}
N=\frac{ V}{(2\pi)^{3} } \frac{ 4 \pi}{3 } k^3_F
 \end{equation}
In the result the total energy of the neutrino ball is
 \begin{equation}
{\cal E}= A \frac{ N^{\frac{ 4}{3 }}}{R } +4\pi \sigma R^2
 \end{equation}
where
\begin{equation}
\sigma=2\sqrt{\lambda'}v^3 I
\end{equation}
with
\begin{equation}
I = \int^{\infty}_{0} dy \frac{e^{2y}}{(1+e^{y})^{4}} =0.0832
\end{equation}
and
\begin{equation}
A= \frac{3}{8}(9\pi )^{\frac{1}{3}}
\end{equation}
$ \sigma$ has an interpretation of the surface tension. It comes from the
thin-shell which bounds the interior of this soliton star and
separates the false and true vacuum.
{}From the expression for the  energy of the soliton it is clear that in this
case
the soliton is stabilized by the surface tension $\sigma$.
The ball radius $R$ can be estimated by minimizing the total energy $\cal E$
the obtained radius and mass are
\begin{equation}
R = (\frac{A}{8\pi \sigma})^{\frac{1}{3}}N^{\frac{4}{9}}
\end{equation}
\begin{equation}
M = \frac{3}{2}A^{\frac{2}{3}}(8\pi \sigma)^{\frac{1}{3}}N^{\frac{8}{9}}
\end{equation}
As the
exponent $N < 1$ for large $N$, the soliton mass is always less then
that of the free particles solution and this ensures its  stability.
However, in the
limiting case of a large number of fermions ($N \gg 1$) the effects of gravity
become important
and can be taken into consideration either by coupling the Newtonian
gravitational
field to the energy density inside the ball or by calculating the
gravitational radius of the soliton.
The calculation of the soliton radius $R$ presented above is equivalent to
other method in which
we can compare the fermions pressure
\begin{equation}
P=-\frac{ \partial {\cal E}}{\partial V }=\frac{ 1}{3 } \rho (r)
\end{equation}
with
 \begin{equation}
\rho=\frac{ 3AN^{4/3}}{4 \pi R^4 }
 \end{equation}
with the surface pressure
 \begin{equation}
P_{\Sigma }= - \frac{ 2 \sigma}{ R }
 \end{equation}
Including gravity the local equilibrium condition
demands
 \begin{equation}
\frac{ 1}{r^2 } \frac{ d}{dr }(\frac{ r^2}{\rho (r) }\frac{ dp}{dr })=
-4\pi G_N \rho (r)
 \end{equation}
Inside the ball the pressure $ P=\rho /3$ so we have the equation
 \begin{equation}
\frac{ 1}{r^2 } \frac{ d}{dr }(\frac{ r^2}{\rho (r) }\frac{ d \rho}{dr })=
-12\pi G_N \rho (r)
 \end{equation}
with the surface boundary condition
 \begin{equation}
P_{\Sigma}(R) +\frac{ 1}{3 } \rho (R) =0
 \end{equation}
Defining
 \begin{equation}
\rho (r)=\rho_0 e^{\varphi}
 \end{equation}
with $ \kappa = 12 \pi G_N \rho_0$ we have the Liuoville equation
 \begin{equation}
\triangle \varphi = -\kappa e^{\varphi}
 \end{equation}
The Laplace operator is
 \begin{equation}
\triangle = \frac{ 1}{r^2 }\frac{ d}{dr } (r^2 \frac{ d}{dr } ) + ...
 \end{equation}
Using the thin-wall approximation one can obtain the following expression
\begin{equation}
e^{\phi} = \frac{1}{ch^{2}(\frac{\sqrt{2}x}{2})}
\end{equation}
where the new variable $x$ is define in the following way
\begin{equation}
r=r_{0}x
\end{equation}
\begin{equation}
r_{0} = \frac{1}{\sqrt{12\pi G_{N}\rho_{0}}}
\end{equation}
This allows to describe the profile of the energy density in the star Fig.1
\begin{equation}
\rho (r) = \frac{\rho_{0}}{ch^{2}(\frac{\sqrt{2}r}{2r_{0}})}.
\end{equation}
For the neutrino-ball surface defined by $r = R$ the density equals
\begin{equation}
\rho (R) = \frac{\rho_{0}}{ch^{2}(\frac{\sqrt{2}R}{2r_{0}})}
\label{wz_6}
\end{equation}
On the other hand considering the case of static spherical ball, for
the equation of state
\begin{equation}
P(r) = \frac{1}{3} \rho (r)
\label{wz_0}
\end{equation}
the internal neutrino pressure
must be balanced by the pressure originated from the surface tension $\sigma$.
This gives the value of the $\rho(R) = \frac{6\sigma}{R}$.
Comparing the last equation with the equation [\ref{wz_6}] we can calculate
the parameter $\rho_{0}$
\begin{equation}
\rho_{0} = \frac{6\sigma}{R}ch^{2}(\frac{\sqrt{2}R}{2r_{0}})
\end{equation}
{}From previous calculations $\rho_{0}$ is given by the relation
\begin{equation}
\rho_{0} = \frac{1}{12\pi G_{N}r_{0}^{2}}
\end{equation}
so we can obtain the equation
\begin{equation}
ux = chx
\label{wz_1}
\end{equation}
where $u$ is given by
\begin{equation}
u = \frac{1}{6\sqrt{\pi \sigma G_{N}R}}
\label{wz_2}
\end{equation}
The equation [\ref{wz_1}] can be presented in the
following way
\begin{equation}
u = f(x)
\end{equation}
where
\begin{equation}
f(x) = \frac{chx}{x}
\end{equation}
Number of roots of this equation depends on the value of minimum of the
function $f(x)$ (Fig.2). If $x_{0}$ represents the point at which $f(x)$
reaches the
minimum then the relation $u = f(x_{0})$ allows
 us to estimate the point
for which exists only one solution of the equation [\ref{wz_1}]. For $f(x) >
f(x_{0})$
the equation [\ref{wz_1}] possesses two roots.
As the next step one should also take into account the Schwarzschild
criterion for the
stability of this star. We are interested only in the situation when $R >
R_{g}$
($R_{g}$  is the gravitational radius of the star).
This condition is achieved for each value of $R$.
The equations [\ref{wz_1}] and [\ref{wz_2}] allows to estimate the mass
of the neutrino ball which is given by the dependence
\begin{equation}
M(R) = 4\pi \sigma R^2 + 4\pi
\frac{6\sigma}{R}ch^2(\frac{R}{\sqrt{2}r_{0}})\int_{0}^{R}\frac{drr^3}{ch^2(\frac{r}{\sqrt{2}})}
\end{equation}
The result namely the mass-radius curve $M(R)$ is presented on Fig 3.
In particular, this soliton model based on the equation of state [\ref{wz_0}]
gives
the following values of parameters $R_{max} = 8.52 \times 10^{15} cm$ and
$M_{max} = 4.58 \times 10^{9} M_{\odot}$ which are consistent with
observational data.

\newpage
\section{Conclusion}
The current idea of quasar is that its energy comes from the  matter
accretion
on the
supermassive black holes. Nevertheless this model can not solve many
astrophysical problems associated for example with the early formation of such
massive black holes.
Alternative explanation of a quasar is connected with the
phenomena of  phase transitions in the early universe.
The grand unification theory predicts the sequence of  phase transitions
during the evolution of the early universe. If they are discontinue then
the bubbles of the new low temperature phase will appear during the
universe expansion. After phase transition point the low temperature phase
will dominate and the areas of the old high temperature phase also will
form bubbles. In the presence of fermions inside, the soliton is stabilized by
surface tension term ($ \sim R^2 $).
As the result the equilibrium
configuration appears with the definite mass and radius.  The comparatively
late phase transition takes place in the standard model during the
spontaneous  symmetry breaking from $ SU_L (2) \times U_Y (1) $ to $U_Q (1)$.
If such a phase transition is discontinue then the bubbles of the high
temperature phase filled  for example  with neutrinos may be produced.
In this paper it was shown that inside the bubble we have only the Dirac
neutrino
with mass of the order of the electron or quark mass. This implies that the
total lepton number is conserved inside the ball.
In the broken phase it was obtained two Majorana mass eigenstates.
In this model ($V \sim 1 \ TeV$) so the biggest drops can reach the size $ R =
8.52 \times 10^{15}$ cm
and the mass $ M = 4.58 \ 10^9 M_{ \odot} $.
If we put such a bubbles into the interstellar medium they may produce the
identical
accretion as we expect from the supermassive black holes.

$$
\pscaption{\boxit{\psboxto(10cm;10cm){fig3.ps}}}
{Fig 1. The radius dependence $(x=r/r_0 )$ of the star density r.}
$$

$$
\pscaption{\boxit{\psboxto(10cm;10cm){fig2.ps}}}
{Fig 2. The x dependence of the function  ch(x)/x}
$$
$$
\pscaption{\boxit{\psboxto(10cm;10cm){fig1.ps}}}
{Fig 3. The mass-radius dependence for the neutrino ball.}
$$

\autojoin
\end{document}